\documentclass[journal]{IEEEtran}
\usepackage{cite}

\ifCLASSINFOpdf
\else
\fi
\usepackage{amsmath}

\usepackage{amsfonts,amssymb,amsmath,url,graphics,subfigure,cite,calc,theorem}
\usepackage{amsmath}

\usepackage{amssymb}
\usepackage{graphicx}
\usepackage{float}

 \usepackage{geometry}
 
\begin{document}

\thispagestyle{empty} 
\noindent
\begin{center}
    \large\textbf{Copyright Notice}
\end{center}

\vspace{2cm}

\noindent
\copyright\ 2016 IEEE. Personal use of this material is permitted. Permission from IEEE must be obtained for all other uses, in any current or future media, including reprinting/republishing this material for advertising or promotional purposes, creating new collective works, for resale or redistribution to servers or lists, or reuse of any copyrighted component of this work in other works.

\vspace{1cm}

\noindent
\textbf{To cite this article:} \\
S Lin, BT Vo, SE Nordholm ``Measurement driven birth model for the generalized labeled multi-Bernoulli filter,'' in \textit{2016 International Conference on Control, Automation and Information Sciences (ICCAIS)}, pp. 94-99, 2016.

\vspace{1cm}

\noindent
\textbf{Official Version of Record:} \\
The final version of this article is available at: 

\url{https://ieeexplore.ieee.org/abstract/document/7822442}

\url{10.1109/ICCAIS.2016.7822442}

\vspace{2cm}
\noindent
\textbf{Dr. Shoufeng Lin has been a Senior Member of IEEE since 2020.}

\newpage 
\setcounter{page}{1} 

 \pagenumbering{gobble}
%
\title{Measurement Driven Birth Model for the Generalized Labeled Multi-Bernoulli Filter}
%
%
%

\author{Shoufeng~Lin,~\IEEEmembership{Member,~IEEE,\thanks{Corresponding author: shoufeng.lin@postgrad.curtin.edu.au; ee.linsf@gmail.com. }}
        ~Ba~Tuong~Vo,~and~Sven~E.~Nordholm,~
        \IEEEmembership{Senior~Member,~IEEE}
}

\maketitle

\begin{abstract}
This paper presents a measurement driven birth (MDB) model for the generalized labeled multi-Bernoulli (GLMB) filter. The MDB model adaptively generates target births based on measurement data, thereby eliminating the dependence of \textit{a priori} knowledge of target birth distributions. Numerical results are provided to demonstrate the performance.
\end{abstract}


%
\IEEEpeerreviewmaketitle

\section{Introduction}
%
%
%
%
\IEEEPARstart{M}{ulti-target} tracking filters have been proposed aiming at jointly estimating an unknown and time-varying number of targets and their individual states from a sequence of observations with detection uncertainty, association uncertainty and clutter. Besides the Multiple Hypotheses Tracking (MHT) and the Joint Probabilistic Data Association (JPDA), the finite set statistics (FISST) forms a new framework that models the multi-target state as an random finite set (RFS), and propagates the state density via the multi-target Bayes recursion \cite{Mahler07}. 

Lately, the Labeled Multi-Bernoulli (LMB) filter and Generalized Labeled Multi-Bernoulli (GLMB) filters have been proposed \cite{LMB, GLMB1,GLMB2} with improved performance including the accuracy and the ability in identifying the trajectory of each target, following the development of Probability Hypothesis Density (PHD), Cardinalized Probability Hypothesis Density (CPHD), and the Cardinality-Balanced Multi-Bernoulli filters \cite{MahlerPHD2,MahlerCPHDAES,VSD05,VVC07,VVC09}.

Standard implementations of the GLMB filters require \textit{a priori} knowledge of target birth distributions, and therefore can be restrictive in practical applications. In this paper, we present an measurement-driven birth distribution model that relies only on measurement data. 

The paper is organized as follows. The necessary background on labeled RFSs, Bayes recursion and GLMB is given in Section \ref{Background}. Section \ref{MDB} presents the MDB model for the GLMB filter. Numerical results are provided in Section \ref{NumericalResults}, and closing remarks are given in Section \ref{conclusion}.

\section{Background} \label{Background}
\subsection{Labeled RFS and Definitions}
According to \cite{GLMB1,GLMB2,LMB}, an RFS is a finite-set-valued random variable. Its number of points is random and the points are random and unordered. The labeled RFS is introduced to accommodate target identity, i.e. each target state $x \in \mathbb{X}$ is uniquely identified by a label $\ell$, where $\mathbb{X}$ is a state space, $\ell \in \mathbb{L}$, $\mathbb{L} = \{ \ell_i : i \in \mathbb{N} \}$, $\mathbb{N}$ denotes the set of positive integers. The resulting labeled RFS with state space $\mathbb{X}$ and discrete label space $\mathbb{L}$, is an RFS on $\mathbb{X} \times \mathbb{L}$, such that each realization has distinct labels.

Throughout the paper, single-target states are represented by lowercase
letters (e.g. $x$, $\mathbf{x}$), while multi-target states are represented by
uppercase letters (e.g. $X$, $\mathbf{X}$), labeled states and their distributions use bold face letters (e.g. $\mathbf{x}$, $\mathbf{X}$, $\mathbf{\pi }$, etc.) to distinguish them from unlabeled ones, spaces are denoted by blackboard bold letters (e.g. $\mathbb{X}$, $\mathbb{Z}$, $\mathbb{L}$, $\mathbb{N}$, etc.), and the class of finite subsets of a space $\mathbb{X}$ is denoted by $\mathbf{\mathcal{F}(}\mathbb{X)}$. 

We use the standard inner product notation
\begin{equation*}
\left\langle f,g\right\rangle \triangleq \int f(x)g(x)dx
\end{equation*}%
and the multi-object exponential notation
\begin{equation}
h^{X} \triangleq \prod \nolimits _{x\in X} h(x)  \label{eq:multiobject_exp}
\end{equation}%
where $h$ is a real-valued function, with $h^{\emptyset }=1$ by convention.

We denote a generalization of the delta function that takes arbitrary arguments such as sets, vectors, integers etc., by
\begin{equation*}
\delta _{Y}(X)\triangleq \left\{
\begin{array}{l}
1,\text{ if }X=Y \\
0,\text{ otherwise}%
\end{array}%
\right. 
\end{equation*}%
and the inclusion function, a generalization of the indicator function, by%
\begin{equation}
1_{Y}(X)\triangleq \left\{
\begin{array}{l}
1,\text{ if }X\subseteq Y \\
0,\text{ otherwise}%
\end{array}%
\right. 
\end{equation}%

Projection $\mathcal{L}:\mathbb{X}\mathcal{\times }\mathbb{L}\rightarrow \mathbb{L}$
is defined as $\mathcal{L}((x,\ell ))=\ell $. Then a finite subset $%
\mathbf{X}$ of $\mathbb{X}\mathcal{\times }\mathbb{L}$ has distinct labels
if and only if $\delta _{|\mathbf{X}|}(|\mathcal{L(}\mathbf{X})|)=1$. Here $|\cdot|$ means cardinality of a set, $\mathcal{L}(\mathbf{X})=\{%
\mathcal{L}(\mathbf{x})\!:\!\mathbf{x}\!\in \!\mathbf{X}\}$. Hence the function $\Delta (\mathbf{X})\triangleq $ $\delta _{|\mathbf{X}|}(|\mathcal{%
L(}\mathbf{X})|)$ is called the \emph{distinct label indicator}.

%

\subsection{Bayes Multi-target Recursion}
Suppose that at time $k$, we have the multi-target state and multi-target observation, respectively $ \mathbf{X}_k = \{ \mathbf{x}_{k,1}, ... , \mathbf{x}_{k,N(k)} \} $, $ {Z}_k = \{ {z}_{k,1}, ... , {z}_{k,M(k)} \}$, where $N(k)$ denotes the number target states and $M(k)$ the number of observations. 

Let $\pi_k(\cdot|{Z}_k)$ denote the multi-target posterior density at time $k$, and $\pi_{k+1|k}$ denote the multi-target prediction density to time $k+1$. The multi-target Bayes recursion involves the update and the prediction steps. 
\begin{align}
\!\!\mathbf{\pi }_{k}(\mathbf{X}_{k}|Z_{k})& =\!\frac{g_{k}(Z_{k}|\mathbf{X}%
_{k})\mathbf{\pi }_{k|k-1}(\mathbf{X}_{k})}{\int g_{k}(Z_{k}|\mathbf{X})%
\mathbf{\pi }_{k|k-1}(\mathbf{X})\delta \mathbf{X}}  \label{eq:MTBayesPred}
\\
\!\!\mathbf{\pi }_{\!k+1|k\!}(\mathbf{X}_{k+1})& =\!\int \!\mathbf{f}%
_{\!k+1|k\!}(\mathbf{X}_{k+1}|\mathbf{X}_{k})\mathbf{\pi }_{k\!}(\mathbf{X}%
_{k}|Z_{k})\delta \!\mathbf{X}_{k}  \label{eq:MTBayesUpdate}
\end{align}%
where $g_k(\cdot|\cdot)$ is the \emph{multi-target likelihood function} at time $k$, $\mathbf{f}_{k+1|k\!}$ is the \emph{multi-target state transition density} to time $k+1$,
and the integral is a \emph{set integral} defined for any function $%
\mathbf{f:\mathcal{F}(}\mathbb{X}\mathcal{\times }\mathbb{L)}\rightarrow
\mathbb{R}$ by%
\begin{equation}
\int \mathbf{f}(\mathbf{X})\delta \mathbf{X}=\sum_{i=0}^{\infty }\frac{1}{i!}%
\int \mathbf{f}(\{\mathbf{x}_{1},...,\mathbf{x}_{i}\})d(\mathbf{x}_{1},...,%
\mathbf{x}_{i}) \\
\end{equation}

A number of multi-target distributions have been proposed to model the unlabeled multi-target density and make the Bayes resursion (\ref{eq:MTBayesPred}, \ref{eq:MTBayesUpdate}) tractable \cite{MahlerPHD2,MahlerCPHDAES,VSD05,VVC07,VVC09}. GLMB is a new model that accommodates not only the multi-target states (e.g. locations) but also the multi-target identities (i.e. labels) in the recursion.

Specifically, for the multi-target labeled RFS, we have
\begin{equation}
\begin{aligned}
 & \int\mathbf{f}(\{\mathbf{x}_{1},...,\mathbf{x}_{i}\}) d(\mathbf{x}_{1},...,%
\mathbf{x}_{i})=  \\
& \sum_{(\ell _{1},...,\ell _{i})\in \mathbb{L}^{i}}\!\int_{\mathbb{X}%
^{i}} \!\! f(\{(x_{1},\ell _{1}),...,(x_{i},\ell _{i})\})d(x_{1},...,x_{i})
\end{aligned}%
\end{equation}

\subsection{GLMB Recursion}
\label{GLMB}
A GLMB RFS is a labeled RFS with state space $\mathbb{X}$ and label space $\mathbb{L}$ with probability density given by (\ref{eq:GLMBRFS}). It can be regarded as a mixture of multi-target exponentials \cite%
{GLMB1}.
\begin{equation} \label{eq:GLMBRFS}
\mathbf{\pi }(\mathbf{X})=\Delta (\mathbf{X})\sum_{\xi \in \Xi }w^{(\xi )}(%
\mathcal{L(}\mathbf{X}))\left[ p^{(\xi )}\right] ^{\mathbf{X}}
\end{equation}%
where $\Xi $ is a discrete index space, each $p^{(\xi )}(\cdot ,\ell )$ is the probability density of the states of target $\ell \in I=\mathcal{L(}\mathbf{X})$, and each $w^{(\xi )}(I)$ is non-negative with $%
\sum_{(I,\xi )\in \mathcal{F}\!(\mathbb{L})\!\times \!\Xi }w^{(\xi )}(I)=1$. 

A Labeled Multi-Bernoulli (LMB) RFS is a special case of a GLMB RFS with a single component:
\begin{equation}
\mathbf{\pi }(\mathbf{X})=\Delta (\mathbf{X})w(%
\mathcal{L(}\mathbf{X})) p ^{\mathbf{X}} \label{eq:generativeLMB}
\end{equation}

To facilitate numerical implementation of GLMB, an alternative form, known as the $\delta$-GLMB, has been proposed as (\ref{eq:generativeGLMB}). 
\begin{equation}
\mathbf{\pi }(\mathbf{X})=\Delta (\mathbf{X})
\!\!\!\! \sum_{(I,\xi )\in \mathcal{F}(%
\mathbb{L})\times \Xi }\omega ^{(I,\xi )}\delta _{I}(\mathcal{L(}\mathbf{X}))%
\left[ p^{(\xi )}\right] ^{\mathbf{X}}  \label{eq:generativeGLMB}
\end{equation}%
where $\omega ^{(I,\xi )}=w^{(\xi )}(I)$. It can be
obtained from the GLMB based on the fact that $w^{(\xi )}(J)=\sum_{I\in
\mathcal{F}\!(\mathbb{L})}w^{(\xi )}(I)\delta _{I}(J)$, since the summand is
non-zero if and only if $I=J$, where $J \in \mathcal{F}(\mathbb{L})$ is a set of labels. 
A $\delta $-GLMB is completely characterized by the set of parameters $\{(\omega ^{(I,\xi )},p^{(\xi )}):(I,\xi )\in \mathcal{F}\!(\mathbb{L}%
)\!\times \!\Xi \}$.

In practice, the probability densities of $\delta $-GLMB are conditioned on measurements up to time $k\geq 0$, and the discrete space $\Xi $ is the space of association map histories $\Theta_{0:k}\triangleq \Theta _{0}\times ...\times \Theta _{k}$, where $\Theta _{t}$ denotes the association map space at time $t$. Here an association map records the association between targets and measurements, i.e. undetected targets are assigned with $0$ at the end of the current association map, while a target $\ell $ that generates a measurement $z_{\theta (\ell )}\in Z$ is assigned with ${\theta (\ell )}$. 

Each $\xi =(\theta _{0},...,\theta _{k})\in \Theta _{0:k}$
represents a history of association map up to time $k$, which also contains the history of target labels encapsulating both births and deaths. A target can generate at most one measurement at any point of time. Similar to the definition of $\Theta
_{0:k}$, $\mathbb{L}_{0:k}$ is the space of target label histories up to time $k$. Hence $I\in \mathcal{F}(\mathbb{L}_{0:k})$ represents a set of target labels at time $k$. For convenience, in the rest of the paper, we do not refer explicitly to time indices unless where necessary. Thus $\mathbb{L\triangleq L}_{0:k}$, $\mathbb{B\triangleq L}%
_{k+1}$, $\mathbb{L}_{+}\mathbb{\triangleq L}\cup \mathbb{B}$, $\mathbf{\pi
\mathbb{\triangleq }\pi }_{k}$, $\mathbf{\pi }_{+}\mathbb{\triangleq }%
\mathbf{\pi }_{k+1|k}$, $p \triangleq p_{k}$, $g\mathbb{\triangleq }g_{k},{f}\mathbb{%
\triangleq }{f}_{\!k+1|k}$, $p_+ \triangleq p_{k+1|k}$, $\omega_+ \triangleq \omega_{k+1|k}$, and $\mathbf{X}_+ \triangleq \mathbf{X}_{k+1|k}$.

\subsubsection{GLMB Update}
If the current multi-target prediction density is a $%
\delta $-GLMB of the form (\ref{eq:generativeGLMB}), then the multi-target
posterior density is a $\delta $-GLMB given by%
\allowdisplaybreaks
\begin{equation}
\begin{aligned} 
 & \mathbf{\pi }\!(\mathbf{X}|Z)= \\ &
\Delta \!(\mathbf{X})\!\!\!\!\!\!\!\!\sum_{(I,%
\xi )\in \mathcal{F}\!(\mathbb{L})\!\times \!\Xi }\;\sum\limits_{\theta
\!\in \Theta \!(I)}\!\!\!\!\omega^{\!(I,\xi ,\theta \!)\!}(Z)\delta
_{\!I\!}(\mathcal{L\!(}\mathbf{X})\!)\!\!\left[ p^{\!(\xi ,\theta )\!}(\cdot
|Z)\right] ^{\!\mathbf{X}}  \label{eq:PropBayes_strong0}
\end{aligned}%
\end{equation}
%
%
where $\Theta (I)$ denotes the subset of current association maps with
domain $I$,\allowdisplaybreaks%
\begin{eqnarray}
\omega ^{(I,\xi ,\theta )\!}(Z)\!\!\! &\propto &\!\!\!\omega ^{(I,\xi
)}[\eta _{Z}^{(\xi ,\theta )}]^{I}  \label{eq:PropBayes_strong1} \\
\eta _{Z}^{(\xi ,\theta )}(\ell )\!\!\! &=&\!\!\!\left\langle p^{(\xi
)}(\cdot ,\ell ),\psi _{Z}(\cdot ,\ell ;\theta )\right\rangle 
\label{eq:PropBayes_strong2} \\
p^{\!(\xi ,\theta )\!}(x,\ell |Z)\!\!\! &=&\!\!\!\frac{p^{(\xi )}(x,\ell
)\psi _{Z}(x,\ell ;\theta )}{\eta _{Z}^{(\xi ,\theta )}(\ell )}
\label{eq:PropBayes_strong3} \\
\psi _{Z}(x,\ell ;\theta )&=&\!\!\!\!\! \left\{
\begin{array}{ll}
\!\!\!\! \frac{p_{D}(x,\ell )g(z_{\theta (\ell )}|x,\ell )}{\kappa (z_{\theta (\ell
)})}, \text{if }\theta (\ell )>0 \\
1-p_{D}(x,\ell ), \text{if }\theta (\ell )=0%
\end{array}%
\right.  \label{eq:PropConj5} 
\end{eqnarray}
$g(z_{\theta (\ell )}|x,\ell )$ is the single target likelihood for the measurement $z_{\theta (\ell )}$ being generated by $(x,\ell)$, and $\kappa(\cdot)$ is the intensity function of Poisson RFS which we use to describe the clutter. $p_D$ is the probability of a target state being detected. 

\subsubsection{GLMB Prediction}
If the current multi-target filtering density is a $%
\delta $-GLMB of the form (\ref{eq:generativeGLMB}), then the multi-target
prediction to the next time is a $\delta $-GLMB given by%
\begin{equation}
\begin{aligned}
\mathbf{\pi }_{\! +} & (\mathbf{X}_{\!+\!})
=  \\ & \Delta(\mathbf{X}%
_{\!+})\!\!\!\!\!\!\!\sum_{(I_{+},\xi )\in \mathcal{F}(\mathbb{L}_{+})\times
\Xi }\!\!\!\!\omega _{+}^{(I_{+},\xi )}\delta _{I_{+\!}}(\mathcal{L(}\mathbf{X}%
_{\!+}))\!\left[ p_{+}^{(\xi )\!}\right] ^{\!\mathbf{X}_{+}}
\label{eq:PropCKstrong1}
\end{aligned}%
\end{equation}
%
%
%
where\allowdisplaybreaks%
\begin{eqnarray}
\!\!\!\omega_+ ^{(I_+,\xi )}\!\! &=&\!\!\omega _{S}^{(\xi )}(I_{+}\cap
\mathbb{L}) w_{B}(I_{+}\cap \mathbb{B})  \label{eq:PropCKstrong2} \\
\!\!\!\omega _{S}^{(\xi )}(L)\!\! &=&\!\![\eta _{S}^{(\xi
)}]^{L}\sum_{I\supseteq L}[1-\eta _{S}^{(\xi )}]^{I-L}\omega ^{(I,\xi )}
\label{eq:PropCKstrongws} \\
\!\!\!\eta _{S}^{(\xi )}(\ell )\!\! &=&\!\!\left\langle p_{S}(\cdot ,\ell
),p^{(\xi )}(\cdot ,\ell )\right\rangle   \label{eq:PropCKstrong_eta} \\
\!\!\!p_{+}^{(\xi )}(x,\ell )\!\! &=&\!\!1_{\mathbb{L}}(\ell )p_{S}^{(\xi
)\!}(x,\ell )+1_{\mathbb{B}\!}(\ell )p_{B}(x,\ell )  \label{eq:PropCKstrong3}
\\
\!\!\!p_{S}^{(\xi )}(x,\ell )\!\! &=&\!\!\frac{\left\langle p_{S}(\cdot
,\ell )f(x|\cdot ,\ell ),p^{(\xi )}(\cdot ,\ell )\right\rangle }{\eta
_{S}^{(\xi )}(\ell )}  \label{eq:PropCKstrong4}
\end{eqnarray}
$f(x|\cdot ,\ell )$ is the state transition function. $\mathbb{B}$ is the space of new-born target labels. The set of new-born targets can be represented by an LMB RFS, where $w_{B}$ is the probability of a birth hypothesis of new-born targets 
and $p_B$ is the probability distribution of kinematic states that belong to the birth targets as per (\ref{eq:generativeLMB}). as per (\ref{eq:generativeLMB}). Standard implementation of GLMB filter assumes known birth probability densities and kinematic states. Details of the adaptive measurement-driven birth will be given in Section \ref{MDB} and \ref{NumericalResults}.

In the GLMB recursion, the pair $(I,\xi )\in \mathcal{F}(\mathbb{L})\times \Xi $ is called a
\emph{hypothesis}, and its associated weight $\omega ^{(I,\xi )}$ the probability of the hypothesis. Similarly the pair $(I_+,\xi
)\in \mathcal{F}(\mathbb{L}_{+})\times \Xi $ is called a \emph{%
prediction hypothesis}, with probability $\omega _{+}^{(I_+,\xi )}$. Respectively $p^{(\xi)}(\cdot | \ell)$ and $p_{+}^{(\xi )}(\cdot
,\ell )$ are the posterior and prediction probability distributions of the kinematic state
of target $\ell $ for association map history $\xi $.

It is not tractable to exhaustively compute
all the components first and then discard those with small weights in the GLMB recursion. Truncations via the ranked assignment algorithm and the $K$-shortest path algorithm have been proposed to find and keep components with high weights without having to propagate all the components \cite{GLMB2}.

\section{Measurement Driven Birth} \label{MDB}
The standard implementation of GLMB filter in Section \ref{GLMB} relies on \textit{a priori} knowledge of target birth distributions, which restricts its applications in practice. Here we present the measurement-driven birth model that initiates the kinematic states and existence probabilities of birth targets based on measurement data from previous time, hence adaptively estimates the target tracks online. 

An adaptive birth model for Sequential Monte Carlo (SMC) implementations of PHD and CPHD filters has been proposed in \cite{mdbPHD}. An MDB for SMC-CBMeMBer has been presented in \cite{mdbCBMB}. The adaptive birth distribution for the LMB filter has also been proposed \cite{LMB}. Similarly, here we present details for the measurement-driven birth distributions for the GLMB filter.

Suppose we have current measurements $Z$ that are not associated with any of persistent tracks. They initiate new-born targets at the next time step. The set of new-born targets is a labeled multi-Bernoulli RFS which can be completely characterized by 
$\{r _{B}^{(\ell)}(z),\;%
p_{B} (\cdot,\cdot;z): \ell = \ell_{B}(z) \}_{z \in Z} $
%
where $\ell_{B}(z)$ denotes the label assigned for the non-empty birth target initiated by measurement $z$ with existence probability of $r_{B}(z)$, and $p_{B} (x,\ell;z)$ is the probability density of the corresponding birth target. 

The probability density of the new-born LMB RFS is
\begin{equation}
\mathbf{\pi } _{B} (\mathbf{X}_{+})
=   \Delta(\mathbf{X}%
_{+\!}) w_{B}(\mathcal{L}(\mathbf{X}_{+})) \left[ p_{B}\!\right] ^{\!\mathbf{X}_{+\!}}
\label{eq:lmbBirth}
\end{equation}
where 
\begin{equation}
\label{eq:omegaB}
w _{B}(I) = \prod\limits_{i\in \mathbb{B}}\left( 1-r_{B}^{(i)}\right)
\prod\limits_{\ell \in I}\frac{1_{\mathbb{B}}(\ell )r_{B}^{(\ell )}}{1-r_{B}^{(\ell
)}}
\end{equation}
which leads to $w_{B}$ as in (\ref{eq:PropCKstrong2}).

Meanwhile, the new-born likelihood for each measurement $z \in Z$ can be found by
\begin{equation} \label{eq:rU}
r_{U} (z) = 1 - \sum_{(I,\xi )\in \mathcal{F}\!(\mathbb{L})\!\times \!\Xi } \sum\limits_{\theta\!\in \Theta \!(I)} 1_{z_{\theta}}(z) \omega^{(I,\xi,\theta)}
\end{equation}
where $\omega^{(I,\xi,\theta)}$ is given in (\ref{eq:PropBayes_strong1}), and the inclusion function here indicates if the measurement $z$ has been assigned to a target by any of the updated hypotheses. It can be seen from (\ref{eq:rU}) that, a measurement which has been used in all hypotheses cannot initiate a new-born target ($r_{U} (z) =0$), while for measurements that have not been assigned to any of the targets, the new-born likelihood is 1.

In (\ref{eq:omegaB}), the existence probability of the Bernoulli MDB at the next time that is initiated by a measurement $z \in Z$ depends on its new-born likelihood obtained from current time:
\begin{equation}
r_{B}(z) = \min \Big( r_{B_{\max}},\;   \lambda_{B} \cdot  \frac{r_{U}(z)}{\sum_{\zeta \in Z} r_{U}(\zeta) }  \Big)
\end{equation}
where $\lambda_{B}$ is the expected number of target birth at the next time, and $r_{B_{\max}} \in [0,1]$ is the maximum existence probability of a new-born target to ensure that the resulting $r_{B}(z)$ does not exceed 1 when $\lambda_{B}$ is too large. 

The value of $r_{B_{\max}}$ can be chosen based on the application. In general, a larger value of $r_{B_{\max}}$ produces faster track confirmation but higher incidence of false tracks, while a smaller value of $r_{B_{\max}}$ produces slower track confirmation but lower incidence of false tracks. The mean cardinality of the new-born labeled multi-Bernoulli RFS is given by the sum of existence probabilities 
\begin{equation}
\sum_{\zeta \in Z} r_{B}(\zeta) \leq \lambda_{B}
\end{equation}

For each measurement $z$ that has non-zero new-born likelihood, a new birth of Bernoulli RFS is generated around the measurement, assuming a Gaussian distribution. Detailed implementation is application dependent and an example will be given in Section \ref{NumericalResults}. In this paper, the probability distribution of the states is given in (\ref{eq:MDBdensity}), which is used in (\ref{eq:PropCKstrong3}) for the measurement-driven birth model.
\begin{equation}
\label{eq:MDBdensity}
p_{B}(x,\ell;z) = \sum_{i=1}^{M_b} \frac{1}{M_b} \delta_{x_z^{(i)}}(x), \; z \in Z
\end{equation} 
\begin{equation} \label{eq:gaussBirth}
x_z^{(i)} \sim \mathcal{N}%
\big(x;m_B(z),P_B(z) \big),\; i=1,...,M_b
\end{equation}
%
where $M_b$ denotes the number of generated states for the birth target. $m_B(z)$ is a function that maps from an observation to its corresponding target state where the information can be recovered. $P_B(z)$ is a variance that specifies the distribution of states of the new-born target. Larger values of $P_B(z)$ result in higher error tolerance, while smaller values give better accuracy in general.
%

\section{Numerical Results} \label{NumericalResults}
Performance evaluations for the Sequential Monte Carlo implementation of the MDB-GLMB are provided in this section.

\begin{figure}[tbh]
\centering
\includegraphics[width=0.48\textwidth]{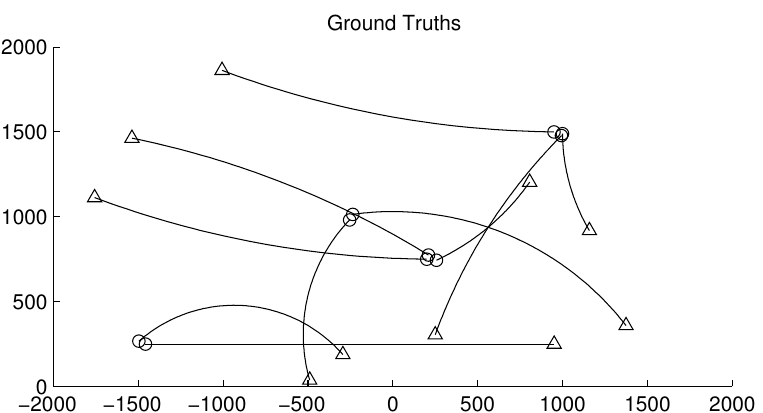}
\centering
\caption{Trajectories of targets. Start/Stop locations for
each target are shown with {\protect\large $\circ $}/{\protect\small $%
\triangle $}.}
\label{fig:xy}
\end{figure}

Consider a non-linear multi-target scenario with 10 targets in total. The
number of targets is time-varying due to births and deaths, and the
observations are subject to missed detections and clutter. Ground truths of targets trajectories are
shown in Figure \ref{fig:xy}. The target state $x=[~\tilde{x}%
^{T},\omega ~]^{T}$ comprises the planar locations and velocity $%
\tilde{x}=[~p_{x},\dot{p}_{x},p_{y},\dot{p}_{y}~]^{T}$ and the
turn rate~$\omega$. Measurements from sensors are of
the form $z=[~\theta,r~]^{T}$ on $x \in [-2000,2000]m$, and $y \in [0, 2000]m$.%

Individual targets follow a coordinated turn model with transition density $%
f(x^{\prime }|x)=\mathcal{N}(x^{\prime };m(x),Q)${,} where \
$m(x)=[~\left[ F(\omega )\tilde{x}\right] ^{T},\omega ~]^{T}$%
, $Q=\mathrm{diag}([\sigma _{w}^{2}GG^{T},\sigma _{u}^{2}])$,
\begin{equation*}
F(\omega )=%
\begin{bmatrix}
1 & \!\!\frac{\sin \omega \Delta }{\omega } & 0 & \!\!-\frac{1-\cos \omega
\Delta }{\omega } \\
0 & \!\!\cos \omega \Delta & 0 & \!\!-\sin \omega \Delta \\
0 & \!\!\frac{1-\cos \omega \Delta }{\omega } & 1 & \!\!\frac{\sin \omega
\Delta }{\omega } \\
0 & \!\!\sin \omega \Delta & 0 & \!\!\cos \omega \Delta%
\end{bmatrix}%
\!\!
\end{equation*}
\begin{equation*}
G=%
\begin{bmatrix}
\frac{\Delta ^{2}}{2} & 0 \\
\Delta & 0 \\
0 & \frac{\Delta ^{2}}{2} \\
0 & \Delta%
\end{bmatrix}%
\!\!,
\end{equation*}%
and $\Delta =1s$ is the sampling time, $\sigma _{w}=5m/s^{2}$ is the
standard deviation of the process noise, $\sigma _{u}=\pi /180rad/s$ is the
standard deviation of the turn rate noise. The survival probability for
targets is $p_{S}(x)=0.99$. 

If detected, each target produces a noisy bearing and range measurement $%
z=[~\theta ,r~]^{T}$ with likelihood $g(z|x)=\mathcal{N}(z;\mu (x),R)$,
where $\mu (x)=[\arctan (p_{x}/p_{y}),\sqrt{p_{x}^{2}+p_{y}^{2}}]^T$ and $R=%
\mathrm{diag}([~\sigma _{\theta }^{2},\sigma _{r}^{2}~])$ with $\sigma
_{\theta }=(2 \pi /180)rad$ and $\sigma _{r}=10m$. The probability of detection
is state dependent and is given by $p_{D}(x)\varpropto \mathcal{N}%
(x;[0,0],\mathrm{diag}([6000,6000])^{2})$, which reaches a peak value of $0.98$ at
the origin and tapers to a value of $0.88$ at the boundary of the observation area. Clutter follows a Poisson RFS with a uniform density on the
observation region with an average of 20 clutter points per scan. 

For the MDB model, we choose $\lambda_{B}=0.3$, $r_{B_{\max}}=0.15$ and $M_b=10000$ here. Each measurement $z$ that initiates a new-born target generates a labeled RFS (with $M_b$ states) around it following a Gaussian distribution (\ref{eq:gaussBirth}), where $m_B(z) = [z(2)\sin(z(1)),0,z(2)\cos(z(1)),0,0]^T$, $P_B(z)=\mathrm{diag}([50,50,50,50,6\pi/180])^2$ is the variance for the new-born states. Each new-born state has a probability density as in (\ref{eq:MDBdensity}).

The OSPA (optimal sub-pattern assignment)\cite{OSPA} metric is used here to evaluate accuracies of the location and the cardinality estimates. The OSPA metric $\bar{d_p}^{(c)}$ of two finite sets $\mathit{X}=\{\mathit{x}_1,...,\mathit{x}_m\}$ and $\mathit{Y}=\{\mathit{y}_1,,...,\mathit{y}_n\} ~ \in \mathcal{F}(\mathbb{X}), (m \leq n) $ is defined as follows.
\begin{equation}
\begin{aligned}
\bar{d_p}^{(c)} & \!(\mathit{X},\!\mathit{Y})\!  \triangleq  \\
& \Big( \frac{1}{n} \big( \min_{\pi \in \Pi_{n}} \sum_{i=1}^{m} d^{(c)}(\mathit{x}_i, \mathit{y}_{\pi(i)})^p + c^p(n-m) \big) \Big)^{\frac{1}{p}}  
\end{aligned}
\end{equation}
where $p \geq 1,~ c > 0$, $ d^{(c)}(\mathit{x},\mathit{y}) \triangleq \min (c, \| \mathit{x} - \mathit{y} \|) , \forall \mathit{x}, \mathit{y} \in \mathbb{X} $, and $\Pi_{n}$ denotes the set of permutations on $\{1,2,...k\},~ k \in \mathbb{N}$. The distance $\bar{d_p}^{(c)}(\mathit{X},\mathit{Y})$ is interpreted as a $p$-th order per-target error. If $m>n,~ \bar{d_p}^{(c)}(\mathit{X},\mathit{Y}) = \bar{d_p}^{(c)}(\mathit{Y},\mathit{X}) $. The order parameter $p$ determines the sensitivity to outliers, and the cut-off parameter $c$ determines the weighting for errors due to cardinality and localization \cite{OSPA}. 

\begin{figure}[tbh]
\centering
\includegraphics[width=0.48\textwidth]{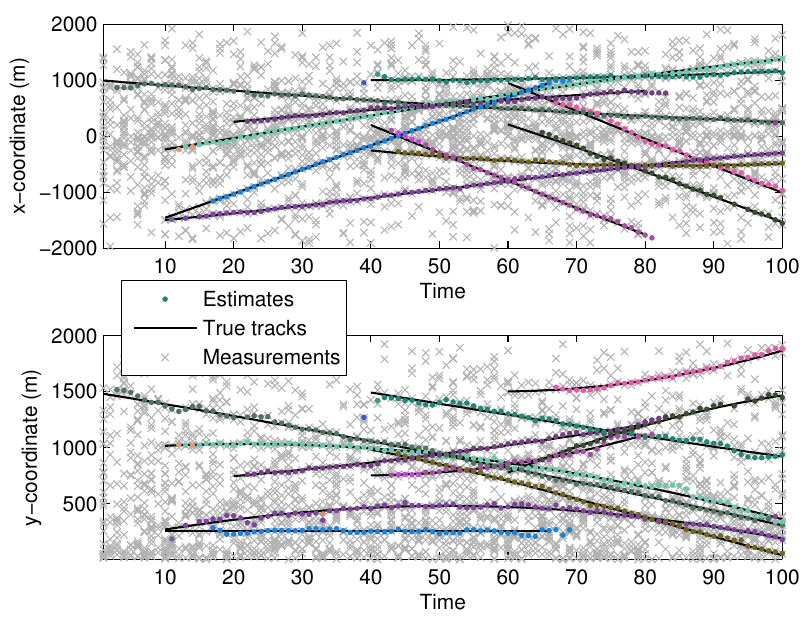}
\centering
\caption{Estimated tracks with identities (in different colors).}
\label{fig:tracks}
\end{figure}

\begin{figure}[t]
\centering
\includegraphics[width=0.48\textwidth]{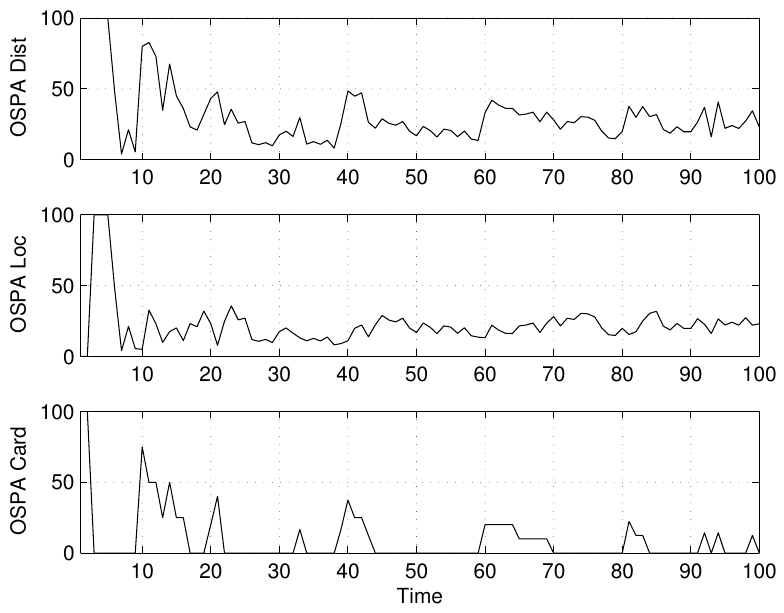}
\centering
\caption{OSPA distance ($c = 100m$, $p=1$).}
\label{fig:ospa}
\end{figure}

\begin{figure}[t]
\centering
\includegraphics[width=0.48\textwidth]{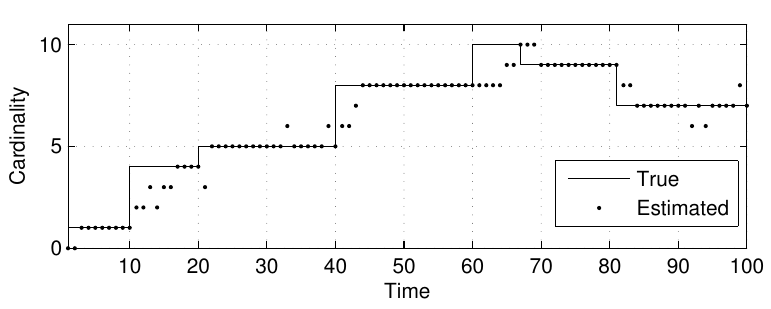}
\centering
\caption{Cardinality estimate.}
\label{fig:card}
\end{figure}

Here we use $c = 100m$, $p=1$. Results are given in Fig.(\ref{fig:tracks},\ref{fig:ospa},\ref{fig:card}). Fig.(\ref{fig:tracks}) presents estimated tracks, Fig.(\ref{fig:ospa}) gives the OSPA performance evaluations \cite{OSPA}, and Fig.(\ref{fig:card}) shows the cardinality estimation results.

\section{Conclusion} \label{conclusion}
This paper presents a measurement-driven birth model for the Generalized Labeled Multi-Bernoulli filter. Results in Section \ref{NumericalResults} show that the MDB-GLMB can track multiple targets by initiating the kinematic states and existence probabilities of birth targets based on measurement data from previous time, and thereby estimating target tracks (with identities) online.

\bibliographystyle{IEEEtran}
\end{document}